\documentclass[11pt,a4paper]{article}

\usepackage{amsmath}

\numberwithin{equation}{section}

\newcommand{\nc}{\newcommand}
\nc{\rnc}{\renewcommand}

\rnc{\title}[1]{\Large\mbox{}\\%\mbox{}\\
     \textbf{#1}\bigskip\medskip\\}
\rnc{\author}[1]{\large #1\\ \smallskip}
\nc{\address}[1]{{\narrower\normalsize\it #1\\}\bigskip}

\nc{\Wt}[4]{W
\begin{pmatrix}
#4 & #3\\
#1 & #2
\end{pmatrix}}

\nc{\Kt}[3]{K
\begin{pmatrix}
#2 &
\begin{matrix}
#3\\
#1
\end{matrix}
\end{pmatrix}}

\nc{\thf}{\vartheta_1}
\nc{\thfr}{\vartheta_4}

\nc{\round}[1]{\left(#1\right)}
\nc{\squareb}[1]{\left[#1\right]}
\nc{\curly}[1]{\left\{#1\right\}}
\nc{\abs}[1]{\left|#1\right|}
\nc{\pointy}[1]{\left\langle #1 \right\rangle}
\nc{\floor}[1]{\left\lfloor #1 \right\rfloor}
\nc{\lsum}[3]{\sum_{#1=#2}^{#3}}
\nc{\smax}{\text{max}}

\nc{\m}{\!-\!}
\nc{\p}{\!+\!}
\nc{\ksum}{\!\!\sum_{k+\frac{1}{2}=1}^\infty\!\!}

\nc{\ks}{\kappa_{\text{s}}}
\nc{\kb}{\kappa_{\text{b}}}
\nc{\als}{\alpha_{\text{s}}}

\nc{\xil}{\xi^{\text{L}}}
\nc{\xir}{\xi^{\text{R}}}
\nc{\xilr}{\xi^{\text{L},\text{R}}}
\nc{\al}{a^{\text{L}}}
\nc{\ar}{a^{\text{R}}}
\nc{\alr}{a^{\text{L},\text{R}}}
\nc{\kslr}{\ks^{\text{L},\text{R}}}
\nc{\epsl}{\eps^{\text{L}}}
\nc{\epsr}{\eps^{\text{R}}}
\nc{\epslr}{\eps^{\text{L},\text{R}}}
\nc{\ksl}{\ks^{\text{L}}}
\nc{\ksr}{\ks^{\text{R}}}

\nc{\lam}{\lambda}
\nc{\Lam}{\Lambda}
\nc{\eps}{\epsilon}
\nc{\veps}{\varepsilon}
\nc{\kap}{\kappa}

\nc{\Dm}{\boldsymbol{D}}
\nc{\Tm}{\boldsymbol{T}}
\nc{\av}{\boldsymbol{a}}
\nc{\bv}{\boldsymbol{b}}
\nc{\Lambold}{\boldsymbol{\Lambda}}
\nc{\Iv}{\boldsymbol{I}}

\DeclareMathOperator{\Tr}{Tr}
\DeclareMathOperator{\Real}{Re}
\DeclareMathOperator{\n}{\mathcal{N}}
\DeclareMathOperator{\sgn}{sgn}
\rnc{\Re}{\Real}

\nc{\sm}[1]{{\scriptstyle #1}}
\nc{\ssm}[1]{{\scriptscriptstyle #1}}
\nc{\spos}[2]{\makebox(0,0)[#1]{$\sm{#2}$}}
\nc{\dl}[3]{\put(#1,#2){\makebox(#3,0){\dotfill}}}
\rnc{\d}[2]{\put(#1,#2){\spos{}{\bullet}}}
\nc{\dd}[3]{\multiput(#1,#2)(0,1){#3}{\spos{}{\bullet}}}

\begin{document}

\begin{center}
\title{Surface Free Energies, Interfacial Tensions\\
and Correlation Lengths of the ABF Models\footnote{Submitted to J. Phys.\ A.}}
\author{
David L. O'Brien\footnote{E-mail address: \texttt{dlo@maths.mu.oz.au}.} and
Paul A. Pearce\footnote{E-mail address: \texttt{pap@maths.mu.oz.au}.  On leave
from the University of Melbourne.}}
\address{\footnotemark[2]Department of Mathematics, The
University of Melbourne,\\ Parkville, Victoria 3052, Australia\\
and\\
\footnotemark[3]Physikalisches Institut der Universit\"{a}t Bonn,\\
Nu\ss allee 12, D-53115 Bonn, Germany.}

\begin{abstract}
\noindent
The surface free energies, interfacial tensions and
correlation lengths of the Andrews-Baxter-Forrester models in regimes~III and
IV are calculated with fixed boundary conditions.
The interfacial tensions are calculated between
arbitrary phases
and are shown to be additive. The associated critical exponents are given by
$2-\als=\mu=\nu$
with $\nu=(L+1)/4$ in regime~III and $4-2\als=\mu=\nu$ with $\nu=(L+1)/2$ in
regime~IV\@. Our results are obtained using general commuting transfer
matrix and inversion relation methods that may be applied to
other solvable lattice models.
\end{abstract}
\end{center}

\section{Introduction}

There has been a recent convergence of interest in statistical mechanics on
systems with a boundary. It is well known that two-dimensional lattice models
without a boundary are exactly solvable~\cite{BaxterBook} by commuting transfer
matrix methods if the local statistical weights satisfy the celebrated
Yang-Baxter equation. It has also been known for some time, from the work of
Cherednik~\cite{Chered} and Sklyanin~\cite{Sk88}, that this integrability
extends to vertex models defined on a strip with open boundaries provided the
local boundary weights satisfy an additional reflection equation or boundary
Yang-Baxter equation. More recently, reflection equations for
interaction-round-a-face (IRF) models have been
introduced~\cite{BPO95,Kul95,AK95}, and integrability has been established for
lattice spin models defined on a strip with fixed or more general boundary
conditions.

Once integrability with a boundary has been established there are various
quantities of physical interest, such as the surface free energies and
interfacial tensions, that one would like to calculate, and methods to achieve
this need to be developed.  In this direction surface free energies of the
Andrews-Baxter-Forrester (ABF)
models in regime~III~\cite{OPB95,ZB95}, the eight-vertex
model~\cite{BZ95}, the dilute $A_L$ models~\cite{BFZ96}, and the CSOS
models~\cite{ZB96} have been obtained by
an extension of the inversion relation method~\cite{Stroganov,Bax82} used to
calculate the bulk free energies.

In this paper we extend the analysis of~\cite{OPB95} to obtain the surface free
energy of the ABF models in regime~IV\@. In addition, we extend the generalized
inversion relation method to calculate also the interfacial tensions and
correlation lengths. In this way, we establish that for solvable lattice models
it is possible to obtain the bulk free energies, the surface free energies, the
interfacial tensions, the correlation lengths and their associated critical
behaviours all by studying the relatively simple inversion relations.

The layout of the paper is as follows. For the rest of this section we follow
reference~\cite{BPO95} in describing the ABF models with fixed boundary
conditions. In sections~2 and 3, we obtain the bulk and surface free energies.
The band structure and ground state degeneracies of the eigenvalue spectra of
the
transfer matrices are discussed in section~4. The calculations of the
interfacial
tensions and correlation lengths are then given in sections~5 and 6
respectively. We conclude with a brief discussion.

\subsection{ABF models with fixed boundaries}

The ABF models \cite{ABF84} are restricted solid-on-solid models in which
heights on the sites of the
square lattice take values in the set $\{1,2,3,\dots,L\}$
subject to the condition that the values of heights on adjacent sites must
differ by $\pm1$.  The Boltzmann weights depend on a
\emph{crossing
parameter} $\lam=\pi/(L+1)$, and a \emph{spectral parameter}~$u$.  Of interest
here are regimes~III and~IV, in which we have $0<u<\lam$. The non-zero face
weights are given by
\begin{align}
\Wt{a}{a\mp1}{a}{a\pm1}&=\frac{\thf(\lam-u)}{\thf(\lam)}\label{wt1}\\
\Wt{a\mp1}{a}{a\pm1}{a}&=
\round{\frac{\thf((a-1)\lam)\thf((a+1)\lam)}{\thf^2(a\lam)}}^{1/2}\,
\frac{\thf(u)}{\thf(\lam)}\\
\Wt{a\pm1}{a}{a\pm1}{a}&=\frac{\thf(a\lam\pm u)}{\thf(a\lam)}.
\end{align}
The $\thf(u)=\thf(u,p)$ is a standard elliptic theta function
with nome $p$.  The temperature variable $t=p^2$ measures the deviation from
criticality.  The critical limit of the ABF models is $t\to0$, approached from
$t>0$ in regime~III and $t<0$ in regime~IV\@.  We therefore express the nome
$p$
in terms of a real parameter $\veps>0$ as
\begin{equation}
p=\begin{cases}
e^{-\pi\veps} & \text{regime III}\\
ie^{-\pi\veps} & \text{regime IV}
\end{cases}
\label{nome}
\end{equation}
so that $t=\pm\exp(-2\pi\veps)$, and $t\to0$ as $\veps\to\infty$.
The product expansions of the functions $\thf$ and $\thfr$ are given by
\begin{gather}
\thf(u,p)=2p^{1/4}\sin u
\prod_{n=1}^\infty (1-2p^{2n}\cos 2u+p^{4n})(1-p^{2n})\\
\thfr(u,p)=\prod_{n=1}^\infty (1-2p^{2n-1}\cos 2u+p^{2(2n-1)})(1-p^{2n}).
\end{gather}
The $\thf$ functions satisfy
the quasiperiodicity properties
\begin{gather}
\thf(u+\pi,p)=-\thf(u,p)\\
\thf(u-i\ln p,p)=-p^{-1}e^{-2iu}\thf(u,p)
\label{quasip}
\end{gather}
and the  ``conjugate modulus'' transformations
\begin{gather}
\thf(u,e^{-\pi\veps})=\frac{1}{\sqrt{\veps}}\,e^{-(u-\pi/2)^2/\pi\veps}\,
E(e^{-2u/\veps},e^{-2\pi/\veps})\label{cmr3}\\
\thf(u,ie^{-\pi\veps})=-\frac{1}{\sqrt{2\veps}}\,e^{i\pi/8}
e^{-(u+\pi/4)^2/\pi\veps}E(e^{u/\veps},-e^{-\pi/2\veps})\label{cmr4}\\
E(x,p)=\prod_{n=1}^\infty(1-p^{n-1}x)(1-p^nx^{-1})(1-p^n).
\end{gather}

Following reference~\cite{BPO95} we introduce boundary
weights. These
weights depend on an additional real
parameter $\xi$, which is independent of $u$ and may be different for the
left and right boundaries.  The non-zero boundary weights are
\begin{equation}
\Kt{a}{a\pm1}{a}=-\sgn(\xi)
\round{\frac{\thf((a\pm1)\lam)}{\thf(a\lam)}}^{1/2}\,
\frac{\thf(u\pm\xi)\thf(u\mp a\lam\mp\xi)}{\thf^2(\lam)}.
\label{boundwts}
\end{equation}
{}From the face weights and boundary weights we construct a
double-row transfer matrix $\Dm(u)$.  For a lattice of width $N$, the entry of
the transfer matrix corresponding to the rows of heights
$\av=\{a_1,\dots,a_{N+1}\}$ and
$\bv=\{b_1,\dots,b_{N+1}\}$ is defined diagrammatically by
\setlength{\unitlength}{12mm}
$$
\pointy{\av|\Dm(u)|\bv}
=\raisebox{-1.4\unitlength}[1.6\unitlength][
1.4\unitlength]{\begin{picture}(6.2,3)(0.4,0)
\multiput(0.5,0.5)(6,0){2}{\line(0,1){2}}
\multiput(1,0.5)(1,0){3}{\line(0,1){2}}
\multiput(5,0.5)(1,0){2}{\line(0,1){2}}
\multiput(1,0.5)(0,1){3}{\line(1,0){5}}
\put(1,1.5){\line(-1,2){0.5}}\put(1,1.5){\line(-1,-2){0.5}}
\put(6,1.5){\line(1,2){0.5}}\put(6,1.5){\line(1,-2){0.5}}
\put(0.5,0.45){\spos{t}{a_1}}\put(1,0.45){\spos{t}{a_1}}
\put(2,0.45){\spos{t}{a_2}}\put(3,0.45){\spos{t}{a_3}}
\put(5,0.45){\spos{t}{a_N}}\put(6,0.45){\spos{t}{a_{N\!+\!1}}}
\put(6.7,0.45){\spos{t}{a_{N\!+\!1}}}
\put(0.5,2.6){\spos{b}{b_1}}\put(1,2.6){\spos{b}{b_1}}
\put(2,2.6){\spos{b}{b_2}}\put(3,2.6){\spos{b}{b_3}}
\put(5,2.6){\spos{b}{b_N}}\put(6,2.6){\spos{b}{b_{N\!+\!1}}}
\put(6.7,2.6){\spos{b}{b_{N\!+\!1}}}
\put(1.05,1.45){\spos{tl}{c_1}}\put(2.05,1.45){\spos{tl}{c_2}}
\put(3.05,1.45){\spos{tl}{c_3}}\put(4.99,1.45){\spos{tr}{c_N}}
\put(5.99,1.45){\spos{tr}{c_{N\!+\!1}}}
\multiput(1.5,1)(1,0){2}{\spos{}{u}}\put(5.5,1){\spos{}{u}}
\multiput(1.5,2)(1,0){2}{\spos{}{\lam\!-\!u}}
\put(5.5,2){\spos{}{\lam\!-\!u}}
\put(0.71,1.5){\spos{}{\lam\!-\!u}}\put(6.29,1.5){\spos{}{u}}
\multiput(0.5,0.5)(0,2){2}{\makebox(0.5,0){\dotfill}}
\multiput(6,0.5)(0,2){2}{\makebox(0.5,0){\dotfill}}
\multiput(1,1.5)(1,0){3}{\spos{}{\bullet}}
\multiput(5,1.5)(1,0){2}{\spos{}{\bullet}}
\end{picture}}
$$
The solid heights $\{c_1,\dots,c_{N+1}\}$ are summed over. As the boundary
weights
are diagonal, we must have $a_1=b_1$ and
$a_{N+1}=b_{N+1}$.  Furthermore, these boundary heights, which we will call
$\al$
and $\ar$, are fixed to the same values for all entries in the transfer matrix.
The parameters $\xil$ and $\xir$ are similarly fixed for all entries.
Defined
in
this way, the double-row transfer matrix exhibits the crossing symmetry
\begin{equation}
\Dm(\lam-u)=\Dm(u)
\label{Dcross}
\end{equation}
and consequently is real symmetric for $u$ real.
More importantly, however, the double-row transfer
matrices form a commuting family,
\begin{equation}
\Dm(u)\Dm(v)=\Dm(v)\Dm(u).
\end{equation}
This implies that the eigenvectors of $\Dm(u)$ are independent of $u$, so that
functional equations satisfied by the transfer matrix are also satisfied by its
eigenvalues.  In particular, all eigenvalues of the transfer matrix satisfy the
crossing symmetry \eqref{Dcross}.
It should be emphasized that all the matrices
in a commuting family share the same boundary heights $\al$ and $\ar$, and the
same
values of $\xil$ and $\xir$.

To ensure that the largest eigenvalue of the double-row transfer matrix is
non-degenerate for all $0<u<\lam$, we impose the restriction
\begin{equation}
\frac{\lam}{2}\le\abs{\xil},\abs{\xir}<\lambda
\label{xirest}
\end{equation}
and in addition require that $\xilr>0$ when $\alr=1$ and $\xilr<0$ when
$\alr=L$ (note that the $\xilr>0$ and $\xilr<0$ regions are disconnected).
Proof of the sufficiency of these restrictions proceeds as follows.

We first show that all the off-diagonal elements of $\Dm(u)$ are non-negative.
Consider an off-diagonal element $\pointy{\av|\Dm(u)|\bv}$, with $\av$ and
$\bv$
such that $\pointy{\av|\Dm(u)|\bv}$ is not identically zero for all $u$.
Since the element is
off-diagonal, we must have $a_i\neq b_i$ for some $2\le i\le N$.  We use the
boundary crossing relation (equation (3.8) of reference~\cite{BPO95}) to
replace
the left boundary weight $K(\lam-u)$ with $K(u)$.  This introduces a face with
spectral parameter $2u-\lam$ and also a (positive) factor of
$\thf(\lam)/\thf(2u)$. We then use the Yang-Baxter equation (equation (3.4)
of reference~\cite{BPO95}) to push this face to the right until it separates
the
$(i-1)$-th and $i$-th pairs of faces.  The upper and lower heights $b_i$ and
$a_i$
differ, so from equation~\eqref{wt1} the weight of this face is
$\thf(2\lam-2u)/\thf(\lam)$, which is positive for $0<u<\lam$.  Since the other
face weights are positive, and since the boundary weights $K(u)$ are positive
for $0<u<\min\curly{\abs{\xil},\abs{\xir}}$, the entry
$\pointy{\av|\Dm(u)|\bv}$
is the sum of positive terms.  The crossing
symmetry~\eqref{Dcross} then dictates that what holds for $u$ must hold for
$\lam-u$, and since $\lam/2\le\min\curly{\abs{\xil},\abs{\xir}}$, the
off-diagonal elements of
$\Dm(u)$ are non-negative for all $0<u<\lam$.

We now observe that the elements of $\Dm(u)$ are bounded, so there exists a
real number $M>0$ such that all elements of the matrix $M\Iv + \Dm(u)$ are
non-negative for $0<u<\lam$.  The largest eigenvalue of this matrix is
non-degenerate by the Perron-Frobenius theorem, and it follows immediately that
the largest eigenvalue of $\Dm(u)$ is also non-degenerate.

Indeed when $u$ satisfies
$\abs{u-\lam/2}\le\min\curly{\abs{\xil},\abs{\xir}}-\lam/2$,
the double-row transfer matrix is
non-negative definite.  This is due to the fact that
when each of the boundary weights $K(u)$ and
$K(\lam-u)$ is non-negative, $\Dm(u)$ is
expressible as the sum of non-negative definite matrices.

When $u=\lam/2$, the symmetry of the face weights is such that the model is
isotropic. In this case the values $\xi=\pm\lam/2$ deserve special
mention since for these choices the isotropic model has all boundary heights
fixed.  This is easily seen from the definition \eqref{boundwts} as, for fixed
$a$, only one of the choices
$a\pm1$ gives a non-zero boundary weight.  The non-zero boundary weights then
contribute only a constant factor to each entry of the transfer matrix.  Aside
from this trivial factor, the lattice exhibits pure fixed boundary conditions,
with boundary heights alternating either $\{a,a+1,a,a+1,\dots\}$ or
$\{a,a-1,a,a-1,\dots\}$.

If we divide each of the face weights by
$\thf(u-\lam/2)$, and each of the boundary weights by $\thf(u-\lam/2)^2$, then
the quasiperiodicity~\eqref{quasip} implies that replacing $u$ by $u-i\ln p$
simply introduces a gauge factor to each of the normalized weights.  These
gauge factors cancel in the entries
of $\Dm(u)$, so it follows that the quasiperiodicity of $\Dm(u)$ is that of
$\thf(u-\lam/2)^{2N+4}$.
If we therefore define a normalized transfer matrix
\begin{equation}
\tilde{\Dm}(u)=\frac{\Dm(u)}{\thf(u-\lambda/2)^{2N+4}}
\end{equation}
then the entries and eigenvalues of $\tilde{\Dm}(u)$ are doubly periodic
functions of $u$ with
\begin{equation}
\text{period rectangle}=\pi\times i\pi\veps^* \qquad
\veps^*=
\begin{cases}
\veps &\text{regime~III}\\
2\veps &\text{regime~IV.}
\end{cases}
\end{equation}
In regime~IV there is an additional symmetry within the period rectangle
\begin{equation}
\tilde{\Dm}(u\pm\pi/2+i\pi\veps)=\tilde{\Dm}(u).
\label{addper}
\end{equation}

As was shown in~\cite{BPO95}, the eigenvalues of the
ABF models with fixed boundary conditions satisfy the inversion identity
\begin{equation}
s_{-1}s_{1}D(u)D(u+\lam)=\epsl\epsr s_{-2}s_{2}f_{-1}f_1 + s^2_0 f_0
D^{1,2}(u)
\label{invid}
\end{equation}
where the functions $s_k$ and $f_k$ are given by
\begin{equation}
s_k=\frac{\thf(2u+k\lam)}{\thf(\lam)} \qquad
f_k=(-1)^N\squareb{\frac{\thf(u+k\lam)}{\thf(\lam)}}^{2N}.
\end{equation}
The functions $\epsl$ and $\epsr$ depend on the left and right
boundary conditions and, for generic $a$ and $\xi$, take the form
\begin{equation}
\epslr=\frac
{\thf(u-\xi)\thf(u+\xi)\thf(u-\xi-a\lambda)\thf(u+\xi+a\lambda)}
{\thf(\lambda)^4}.
\end{equation}
The function $D^{1,2}(u)$ is an eigenvalue of the double-row transfer matrix at
fusion level $1\times 2$.
We observe as in the periodic case
that for large $N$, $f_0D^{1,2}/(f_{-1}f_1)$ is exponentially
small in $-\lam/2<\Re(u)<\lam/2$.  Hence, for $N$ large, the second term in
the inversion identity \eqref{invid} can be neglected. So in
calculating bulk and surface properties we just need to solve the
simple inversion relation
\begin{equation}
s_{-1}s_{1}D(u)D(u+\lam)=\epsl\epsr s_{-2}s_{2}f_{-1}f_1
\label{inversion}
\end{equation}
and the crossing relation
\begin{equation}
D(u)=D(\lambda-u)
\label{evcross}
\end{equation}
subject to the appropriate analyticity and quasiperiodicity in an open strip
containing $0\le\Re(u)\le\lambda$. These properties determine the bulk and
surface quantities uniquely.

The largest eigenvalue $D_{\smax}(u)$ factorizes into contributions
from the bulk, the surfaces and the interface
\begin{equation}
D_{\smax}(u)\sim \kb(u)^{2N}\ks(u)\,\Lambda(u)
\quad\text{as $N\to\infty$}.
\label{sep}
\end{equation}
When the left and right boundary conditions favour the same phase, which is the
case when $n(\alpha,\beta)=0$ in the notation of section~4, there is no
interface and $\Lambda(u)=1$.  When the left and right boundary conditions
favour different phases, which is the case when $n(\alpha,\beta)\ge 1$, there
is
an interface and the factorization applies to all the eigenvalues in the
first band.
Clearly, the inversion and crossing relations factorize into bulk and surface
terms and can be solved sequentially for the bulk free energies,
the surface free
energies ($n(\alpha,\beta)=0$) and the interfacial tensions
($n(\alpha,\beta)\ge 1$). This we do in the following sections.

\section{Bulk free energies}

Equating the bulk terms in the inversion relation \eqref{inversion} implies
that the bulk partition function per site
satisfies the functional equation
\begin{equation}
\kb(u)\kb(u+\lam)=\frac{\thf(\lam-u)\thf(\lam+u)}{\thf(\lam)^2}.
\label{kapinv}
\end{equation}
This equation is of course independent of the boundary conditions on the
lattice and agrees with the equation for periodic boundaries.  It has been
solved previously, but we
include the solution here for completeness.  We use the standard techniques
developed by Stroganov~\cite{Stroganov} and Baxter \cite{Bax82}.

Since the eigenvectors of $\Dm(u)$ are independent of $u$, it follows that the
eigenvalues $D(u)$ have the same analyticity and quasiperiodicity as the
elements of the transfer matrix.
In particular, the eigenvalues must be entire
functions of the spectral parameter $u$. They are therefore completely
characterized by their zeros and growth at infinity.
In regime~III the zeros of the
largest eigenvalue accumulate on the lines
$\Re(u)=-\lam/2$ and $\Re(u)=3\lam/2$.  In addition to these lines, in
regime~IV zeros accumulate on $\Re(u)=(3\lam-\pi)/2$ and
$\Re(u)=(\pi-\lam)/2$ in accordance with the periodicity~\eqref{addper}.
The strip
$-\lam/2<\Real(u)<3\lam/2$ is free of order~$N$ zeros in both regimes. This is
confirmed by  numerical studies for
large finite $N$.  So inside this strip $\kb(u)$ is non-zero and
$\ln\kb(u)$ is analytic.
The quasiperiodicity
\begin{equation}
\kb(u+i\pi\veps^*)=
\begin{cases}
e^{i(\lam-2u)}e^{\pi\veps^*}\kb(u) & \text{regime~III}\\
e^{2i(\lam-2u)}e^{2\pi\veps^*}\kb(u) & \text{regime~IV}
\end{cases}
\label{qper}
\end{equation}
implies that the second derivative of $\ln\kb(u)$ is periodic, with period
$i\pi\veps^*$.  Hence $\ln\kb(u)$ can be expanded in the form of a generalized
Fourier series
\begin{equation}
\ln\kb(u)=Au^2+Bu+\sum_{k=-\infty}^{\infty}c_ke^{2ku/\veps^*}.
\end{equation}
To evaluate the coefficients $c_k$, $A$ and $B$, we take the logarithm in
\eqref{kapinv}, expand the right hand side using the appropriate conjugate
modulus transformations~\eqref{cmr3} and~\eqref{cmr4}, and equate coefficients.

\subsection{Regime III}

In regime~III we have $t>0$, so
we rewrite the right hand side of \eqref{kapinv} using the conjugate
modulus transformation~\eqref{cmr3}.
With both sides of \eqref{kapinv} expanded in
powers of $\exp(2u/\veps)$, we match coefficients and impose the crossing
symmetry \eqref{evcross} to obtain the solution
\begin{equation}
\ln\kb(u)=\frac{1}{\pi\veps}(\lam-u)u +
2\sum_{k=1}^{\infty}
\frac{\cosh[(\pi-2\lam)k/\veps]\sinh[(\lam-u)k/\veps]\sinh(uk/\veps)}
{k\sinh(\pi k/\veps)\cosh(\lam k/\veps)}.
\label{bulkfe}
\end{equation}
Inside the region $-\lam/2<\Real(u)<3\lam/2$ this function gives the bulk
behaviour of the largest eigenvalue of the transfer matrix.  Applying the
Poisson summation formula to the infinite sum gives the behaviour of the free
energy in the critical limit $t\to 0^+$.  When $L$ is even $\ln\kb(u)$ is
regular, but when
$L$ is odd the leading-order singularity is~\cite{ABF84}
\begin{equation}
\ln\kb\sim t^{\pi/2\lam}\ln t.
\end{equation}
Since $\ln\kb\sim t^{2-\alpha}$, when $L$ is odd the bulk critical exponent
$\alpha$ is given by
\begin{equation}
2-\alpha = \frac{\pi}{2\lambda} = \frac{L+1}{2}.
\label{bulkalpha}
\end{equation}

\subsection{Regime IV}

In regime~IV the temperature variable $t$ is negative, so we use the conjugate
modulus transformation~\eqref{cmr4} to rewrite the right hand side
of \eqref{kapinv} in powers of $\exp(u/\veps)$.  As before, we match
coefficients and use the crossing symmetry~\eqref{evcross} to obtain
\begin{multline}
\ln\kb(u)=\frac{1}{\pi\veps}(\lam-u)u
+\sum_{k=1}^\infty
\frac{\cosh[(\pi/2-2\lam)k/\veps]\sinh[(\lam-u)k/\veps]\sinh(uk/\veps)}
{k\sinh(\pi k/2\veps)\cosh(\lam k/\veps)}
\\
\mbox{}+\ksum
\frac{\sinh[(\pi/2-2\lam)k/\veps]\sinh[(\lam-u)k/\veps]\sinh(uk/\veps)}
{k\cosh(\pi k/2\veps)\cosh(\lam k/\veps)}.
\end{multline}
Once again we apply the Poisson summation
formula to obtain the leading order singularity~\cite{ABF84}
\begin{equation}
\ln\kb\sim\begin{cases}
(-t)^{\pi/2\lam} & \text{$L$ even}\\
(-t)^{\pi/2\lam}\ln(-t) & \text{$L$ odd}
\end{cases}
\end{equation}
so that in regime~IV, the critical exponent $\alpha$ is given
by equation~\eqref{bulkalpha} for all $L$.

\section{Surface free energies}

In order to calculate the surface free energies, we assume
that the boundary conditions on the left and the right edges of
the lattice favour the same phase, so that $n(\alpha,\beta)=0$ in the notation
of section~4. In this case there is a unique largest eigenvalue separated from
the other eigenvalues by a gap.  We can then
divide out the bulk quantities in the inversion relation~\eqref{inversion} to
obtain an inversion relation for the surface partition function per site
$\ks(u)$. Explicitly,
\begin{equation}
\ks(u)\ks(u+\lam)=\frac{\thf(2\lam-2u)\thf(2\lam+2u)}
{\thf(\lam-2u)\thf(\lam+2u)}\,\epsl(u)\epsr(u). \label{boundinv}
\end{equation}
The form of this inversion relation suggests a natural factorization of
$\ks(u)$ into a term independent of the boundary conditions, a term dependent
on the left boundary condition, and a term dependent on the right boundary
condition.  We therefore write
\begin{equation}
\ks(u)=\ks^0(u)\ksl(u)\ksr(u).
\end{equation}
The solution of equation~\eqref{boundinv} proceeds in a similar fashion to the
solution of the bulk inversion relation, but whereas the analyticity of
$\ln\kb(u)$ depended on the absence of order~$N$ zeros, the analyticity of
$\ln\ks(u)$ depends on the absence of order~$1$ zeros.  Our numerical studies
show that when $n(\alpha,\beta)=0$
the largest eigenvalue
of the double-row transfer matrix is indeed non-zero for
$0\le\Re(u)\le\lam$.  We therefore conclude that $\ln\ks(u)$ is analytic
on this strip, and along with the quasiperiodicity
\begin{equation}
\ks(u+i\pi\veps^*)=
\begin{cases}
e^{4i(\lam-2u)}e^{4\pi\veps^*}\ks(u) & \text{regime~III}\\
e^{8i(\lam-2u)}e^{8\pi\veps^*}\ks(u) & \text{regime~IV}
\end{cases}
\end{equation}
this allows us to expand $\ln\ks(u)$ as a generalized Fourier series.

\subsection{Regime III}

In regime~III we use once again
the conjugate modulus transformation~\eqref{cmr3} and the crossing
symmetry~\eqref{evcross} to match Fourier coefficients and obtain
\begin{multline}
\ln\ks^0(u)=\\
\frac{\lam}{\pi\veps}(\pi-3\lam)+2\lsum{k}{1}{\infty}
\frac{\sinh[(\pi-3\lam)k/\veps]
\sinh(\lam k/\veps)\cosh[2(\lam-2u)k/\veps]}
{k\sinh(\pi k/\veps)\cosh(2\lam k/\veps)}
\label{bfe}
\end{multline}
and for generic $a$ and $\xi$,
\begin{multline}
\ln\kslr(u)=
\frac{1}{\pi\veps}\squareb{(a\lam+\xi)(\pi-2\xi)+(|\xi|-2\lam)\pi+
(2-a^2)\lam^2+2u(\lam-u)}\\
\mbox{}-2\lsum{k}{1}{\infty}\frac{\cosh[(a\lam+\xi-|\xi|)k/\veps]
\cosh[(\pi-a\lam-\xi-|\xi|)k/\veps]
\cosh[(\lam-2u)k/\veps]}
{k\sinh(\pi k/\veps)\cosh(\lam k/\veps)}\\
\mbox{}+2\lsum{k}{1}{\infty}\frac{\cosh[(\pi-2\lam)k/\veps]}
{k\sinh(\pi k/\veps)}.
\label{xipart}
\end{multline}
We note that, as it should, the height reversal transformation $a\to
L+1-a$ and $\xi\to-\xi$ leaves \eqref{xipart} unchanged.

When the Poisson summation formula is applied to the above expressions for
$\ln\ks^0(u)$ and $\ln\kslr(u)$, the dominant behaviour as $t\to 0^+$ is seen
to come from equation~\eqref{bfe}.  If $L\equiv2\pmod{4}$, however,
this term is regular and we must consider the sum in equation~\eqref{xipart}.
This latter sum is also regular when $\xi>0$ with $a$ odd and when $\xi<0$ with
$a$ even~\cite{OPB95}, but such exceptions aside, the leading-order
singularities of the surface free energy have the form
\begin{equation}
\ln\ks\sim
\begin{cases}
t^{\pi/4\lam} &\text{$L\equiv0$ or $1\pmod{4}$} \\
t^{\pi/2\lam} &\text{$L\equiv2\pmod{4}$} \\
t^{\pi/4\lam}\ln t &\text{$L\equiv3\pmod{4}$.}
\end{cases}
\end{equation}
Since $\ln\ks\sim t^{2-\als}$,
the surface critical exponent $\als$~\cite{Binder,Diehl} in regime~III is
given by
\begin{equation}
2-\als=\begin{cases}
(L+1)/2 & L\equiv 2\pmod{4}\\
(L+1)/4 & \text{otherwise}.
\end{cases}
\end{equation}

\subsection{Regime IV}

In regime~IV we use the conjugate modulus transformation~\eqref{cmr4} to
match coefficients and obtain
\begin{multline}
\ln\ks^0(u) = \frac{\lam}{2\pi\veps}(\pi-6\lam) +
\sum_{k=1}^\infty
\frac{\sinh[(\pi/2-3\lam)k/\veps]\sinh(\lam k/\veps)
\cosh[2(\lam-2u)k/\veps]}
{k\sinh(\pi k/2\veps)\cosh(2\lam k/\veps)}\\
\mbox{}+\ksum
\frac{\cosh[(\pi/2-3\lam)k/\veps]\sinh(\lam k/\veps)
\cosh[2(\lam-2u)k/\veps]}
{k\cosh(\pi k/2\veps)\cosh(2\lam k/\veps)}.
\label{rIVa}
\end{multline}
The form of $\ln\kslr(u)$ depends on the values of $a$ and $\xi$.  When either
$a\le(L-1)/2$, or $L/2\le a\le(L+2)/2$ with $\xi<0$, $\ln\kslr(u)$ is given by
\begin{multline}
\ln\kslr(u)=\frac{1}{2\pi\veps}\squareb{(a\lam+\xi)(\pi-4\xi)+(|\xi|-2\lam)\pi
+2(2-a^2)\lam^2 + 4u(\lam-u)} \\
\mbox{}+\sum_{k=1}^\infty \frac{\cosh[(\pi/2-2\lam)k/\veps]}
{k\sinh(\pi /2\veps)} + \ksum
\frac{\sinh[(\pi/2-2\lam)k/\veps]}{k\cosh(\pi k/2\veps)}\\
\mbox{}-\sum_{k=1}^\infty \frac{\cosh[(a\lam+\xi-|\xi|)k/\veps]
\cosh[(\pi/2-a\lam-\xi-|\xi|)k/\veps]\cosh[(\lam-2u)k/\veps]}
{k\sinh(\pi k/2\veps)\cosh(\lam k/\veps)} \\
\mbox{}-\ksum \frac{\cosh[(a\lam+\xi-|\xi|)k/\veps]
\sinh[(\pi/2-a\lam-\xi-|\xi|)k/\veps]\cosh[(\lam-2u)k/\veps]}
{k\cosh(\pi k/2\veps)\cosh(\lam k/\veps)}.
\end{multline}
On the other hand,
when either $a\ge(L+3)/2$, or $L/2\le a\le(L+2)/2$ with $\xi>0$,  $\ln\kslr(u)$
is given by
\begin{multline}
\ln\kslr(u)=\frac{1}{2\pi\veps}
\squareb{(a\lam\p\xi)(3\pi\m 4\xi)+(|\xi|\m 2\lam\m \pi)\pi
+2(2\m a^2)\lam^2 + 4u(\lam\m u)} \\
\mbox{}+\sum_{k=1}^\infty \frac{\cosh[(\pi/2-2\lam)k/\veps]}
{k\sinh(\pi k/2\veps)} + \ksum
\frac{\sinh[(\pi/2-2\lam)k/\veps]}{k\cosh(\pi k/2\veps)}\\
\mbox{}-\sum_{k=1}^\infty \frac{\cosh[(\pi\m a\lam\m \xi\m |\xi|)k/\veps]
\cosh[(a\lam\m \pi/2\p\xi\m |\xi|)k/\veps]\cosh[(\lam\m 2u)k/\veps]}
{k\sinh(\pi k/2\veps)\cosh(\lam k/\veps)} \\
\mbox{}-\ksum \frac{
\cosh[(\pi\m a\lam\m \xi\m |\xi|)k/\veps]
\sinh[(a\lam\m \pi/2\p\xi\m |\xi|)k/\veps]\cosh[(\lam\m 2u)k/\veps]}
{k\cosh(\pi k/2\veps)\cosh(\lam k/\veps)}.
\end{multline}
We note that once again $\ln\ks(u)$ is unchanged under the height
reversal transformation
$a\to L+1-a$ and $\xi\to-\xi$.

The leading-order singular behaviour of the surface free energy is derived from
the sums in equation~\eqref{rIVa}.  We find
\begin{equation}
\ln\ks\sim\begin{cases}
(-t)^{\pi/4\lam}\ln(-t) &\text{$L\equiv3\pmod{4}$}\\
(-t)^{\pi/4\lam} &\text{otherwise.}
\end{cases}
\end{equation}
Hence in regime~IV we have for all $L$
\begin{equation}
2-\als=\frac{\pi}{4\lam}=\frac{L+1}{4}.
\end{equation}

\section{Band structure and ground state degeneracies}

In regimes~III and IV the ABF models admit a number of coexisting
phases~\cite{ABF84,Huse84}
corresponding to ground state (zero temperature, $|t|\to1$)
configurations on the lattice given by a pair $(a,b)$ such that $a-b=\pm 1$.
Explicitly, the ground states consist of a chequerboard arrangement with height
$a$ on one sublattice and height $b$ on the other
sublattice of the square lattice. At each boundary the
boundary height $a$ together with the choice of $\xi$ determines
the favoured boundary configuration. If $\xi>0$ the favoured boundary
condition is $\{a, a+1, a, a+1,\dots\}$, and if $\xi<0$ the favoured
boundary condition is $\{a, a-1, a, a-1, \dots\}$. Clearly,
the boundary condition can
pick out the phase near the boundaries. When the phase selected on the left
and right boundaries is the same there will be a unique phase and a unique
ground state. However,
when the phase selected by the left boundary differs from that selected
by the right boundary, there will be an interface formed between the two
coexisting phases. In this case there will be an interfacial tension,
the ground state will be highly degenerate corresponding to the many possible
locations of the interface, and the double-row transfer matrix will have a
continuous band of largest eigenvalues in the thermodynamic limit.
More generally, we expect the spectrum of the double-row transfer matrix
to consist of many such continuous bands of eigenvalues. These bands
can overlap with each other but within each band
the states are characterized by a fixed number of domain walls. Furthermore,
the number of eigenvalues in each band can be determined by the simple
combinatorics of counting the compatible ground state configurations.

\subsection{Regime III}

In regime~III there are $2L-2$ coexisting phases. The corresponding
ground state lattice configurations are given by
the pairs $(a,b)$ with $|a-b|=1$, and all values $1\le a,b\le L$ allowed.
Let us denote the left boundary condition with height $a=\al$ by
$a^+$ if $\xil>0$ and $a^-$ if $\xil<0$.
Similarly, for the right boundary with height $b=\ar$ we have $b^+$ or $b^-$
depending on the sign of
$\xir$. This is convenient since under the restriction~\eqref{xirest}
the interfacial tensions and correlation lengths will depend only on the signs
of $\xil$ and $\xir$.
The ground state degeneracies for a double-row transfer matrix of $N$
faces with fixed boundary heights $a$ and $b$ are then given by
\begin{align}
\n(a^+,b^+) &= \n(a^-,b^-) = \binom{(N+|a-b|)/2}{|a-b|} \\
\n(a^+,b^-)&= \binom{(N+|a-b+2|)/2}{|a-b+1|} \label{npm}\\
\n(a^-,b^+)&= \binom{(N+|a-b-2|)/2}{|a-b-1|}
\end{align}
For ease of reference, we define $m(\alpha,\beta)$ and $n(\alpha,\beta)$ by
\begin{equation}
\n(\alpha,\beta)=\binom{m(\alpha,\beta)}{n(\alpha,\beta)}
\end{equation}
where $\alpha=a^\pm$ and $\beta=b^\pm$.
To see how these formulas arise, consider for simplicity an $(a^+,b^-)$
boundary
height configuration with $b-a\ge 2$.  Moving from $a$ to $b$, there must be
$b-a-1$ ground state transitions (domain walls), starting with the state
$\{a,a+1,a,a+1,\dots\}$ and ending with the state $\{\dots,b-1,b,b-1,b\}$.  If
$a$ and $b$ are joined by a path of $N$ steps, and we let
$m_1,m_2,\dots,m_{b-a-1}$ be the steps at which the transitions occur, then the
$m_i$ can take the values
\begin{align}
m_1&=2,4,\dots,N+a-b+2\\
m_i&=m_{i-1}+1,m_{i-1}+3,\dots,N+a-b+i+1\quad\text{for $2\le i\le b-a-1$.}
\end{align}
We rewrite and sum over these combinations using the variables
%We sum over these combinations using the variables
$k_1,k_2,\dots k_{b-a-1}$ such that $m_i=2k_i+i+1$. This gives
\begin{equation}
\n(a^+,b^-)=\sum_{k_1=0}^{\frac{N+a-b}{2}}\sum_{k_2=k_1}^{\frac{N+a-b}{2}}
\dots\sum_{k_{b-a-1}=k_{b-a-2}}^{\frac{N+a-b}{2}}1
=\binom{(N+b-a-2)/2}{b-a-1}
\end{equation}
in agreement with equation~\eqref{npm}.  The other expressions may be derived
from symmetries under the interchange of $a$ and $b$, and from the equivalence
of
$(a^-,\beta)$ with
$N$ steps and
$((a-1)^+,\beta)$ with $N+1$ steps.

\subsection{Regime IV}

First consider $L$ even. In this case there are
$2L-4$ phases, and the ground state configurations are as in regime~III,
with the exclusion of the $(L/2,L/2+1)$ and $(L/2+1,L/2)$ pairs.
In the case when $L$ is odd, there are again
$2L-4$ phases and the ground state lattice configurations are as in
regime~III, except now the pairs $((L-1)/2,(L+1)/2)$ and $((L+1)/2,(L+3)/2)$
are excluded and replaced by the disordered mixture $((L+1)/2,(L+1)/2\pm 1)$
where the heights on one sublattice can take the values $(L-1)/2$ and $(L+3)/2$
independently at each site.

For both even and odd $L$, we find that the number of eigenvalues
in the first band is given as follows ($\floor{x}$ denotes
the largest integer less than or equal to $x$):
\begin{itemize}
\item If $1\le a,b\le \floor{(L+1)/2}$, excluding $\floor{(L+1)/2}^+$, the
degeneracies are the same as in regime III.
\item If $\floor{(L+2)/2}\le a,b\le L$, excluding $\floor{(L+2)/2}^-$, the
degeneracies are the same as in regime III.
\item Otherwise, the degeneracies are
\begin{align}
\n(a^+,b^+) &= \n(a^-,b^-) = \binom{(N+|a-b|-2)/2}{|a-b|-1} \\
\n(a^+,b^-) &= \binom{(N+|a-b+2|-2)/2}{|a-b+1|-1} \\
\n(a^-,b^+) &= \binom{(N+|a-b-2|-2)/2}{|a-b-1|-1}.
\end{align}
\end{itemize}

\section{Interfacial tensions}

The interfacial tension $\sigma^{(\alpha,\beta)}$ between phases
$\alpha=a^\pm$ and $\beta=b^\pm$ is given by
\begin{equation}
-\sigma^{(\alpha,\beta)}=\lim_{P\to\infty}\lim_{N\to\infty}
P^{-1} \ln \Tr \Lambold(u)^{P/2}\label{sigma}
\end{equation}
where
\begin{equation}
\Lambold(u)=\Lambold^{(\alpha,\beta)}(u)=\frac{\Dm^{(\alpha,\beta)}(u)}
{\left[D^{(\alpha,\alpha)}_{\smax}(u)
D^{(\beta,\beta)}_{\smax}(u)\right]^{1/2}}
\end{equation}
and a factor of the inverse temperature has been absorbed into the
definition of $\sigma^{(\alpha,\beta)}$. Here $D^{(\alpha,\alpha)}_{\smax}(u)$
denotes the maximum eigenvalue of the double-row transfer matrix
$\Dm^{(\alpha,\alpha)}(u)$. Note that the bulk and surface contributions
cancel out in the ratio $\Lambold(u)$.
{}From the inversion relation, crossing, and quasiperiodicity, it follows that
the eigenvalues $\Lambda(u)$ satisfy the simple functional equations
\begin{equation}
\Lambda(u)\Lambda(u+\lambda)=1 \qquad \Lambda(u)=\Lambda(\lambda-u)
\label{redinv}
\end{equation}
subject to the periodicity
\begin{equation}
\Lambda(u)=\Lambda(u+i\pi\veps^*)\qquad
\veps^*=
\begin{cases}
\veps &\text{regime~III}\\
2\veps &\text{regime~IV.}
\end{cases}
\label{funcperiod}
\end{equation}

In the thermodynamic limit the eigenvalues of $\Lambold(u)$ form
continuous bands. In this limit the sum in the trace of \eqref{sigma}
is replaced by integrals over these bands. But in the limit of $P$ large
only the first band of largest eigenvalues is expected to contribute to the
interfacial tension. From the counting of degeneracies, this first band
contains $\n(\alpha,\beta)$ eigenvalues.
The largest eigenvalue of $\Dm^{(\alpha,\beta)}(u)$ has zeros which
accumulate on the lines $\Re(u)=-\lambda/2$ and $\Re(u)=3\lambda/2$ in both
regimes, and also on the lines $\Re(u)=(3\lambda-\pi)/2$ and
$\Re(u)=(\pi-\lambda)/2$ in regime~IV\@.
In addition, each eigenvalue in the first band of eigenvalues of
$\Dm^{(\alpha,\beta)}(u)$ has $n(\alpha,\beta)$ pairs of zeros in the
strip $-\lambda/2 < \Re(u) < 3\lambda/2$ at
\begin{equation}
u=\lambda/2\pm i\phi_k \qquad k=1,2,\dots,n(\alpha,\beta)
\end{equation}
where each $\phi_k$ is real and non-zero.

We now seek a solution to the inversion relations~\eqref{redinv} in the
physical strip subject to the given
zeros and periodicity. The required solution is
\begin{equation}
\Lambda(u)=\prod_{k=1}^{n(\alpha,\beta)}\Phi(u+i\phi_k)\Phi(u-i\phi_k)
\end{equation}
where
\begin{equation}
\Phi(u)=\frac
{\thf(\frac{\pi u}{2\lambda}-\frac{\pi}{4},|t|^\nu)}
{\thf(\frac{\pi u}{2\lambda}+\frac{\pi}{4},|t|^\nu)}
\label{PhiIII}
\end{equation}
and
\begin{equation}
\nu=\begin{cases}
(L+1)/4 & \text{regime~III}\\
(L+1)/2 & \text{regime~IV.}
\end{cases}
\end{equation}
Recall that $|t|=|p^2|=\exp(-2\pi\veps)$ where the nome $p$ is defined in
equation~\eqref{nome}.

In the thermodynamic limit the distribution of each $\phi_k$ becomes dense,
yielding
a density $\rho(\phi_1,\dots,\phi_{n(\alpha,\beta)})$. So integrating over
this band of eigenvalues for real values of $u$ in the interval
$0<u<\lambda$ gives
\begin{equation}
\lim_{N\to\infty}\Tr \Lambold(u)^{P/2}\sim
\int_0^{\frac{\pi\veps^*}{2}}\!\!\!\dots\int_0^{\frac{\pi\veps^*}{2}}\!\!\!
\rho(\phi_1,\dots,\phi_{n(\alpha,\beta)})
\!\prod_{k=1}^{n(\alpha,\beta)}\!
\left|\Phi(u+i\phi_k)\right|^P d\phi_1\dots d\phi_{n(\alpha,\beta)}.
\end{equation}
%\clearpage
For $P$ large this multiple integral can be evaluated by saddle point methods.
The saddle point occurs at
\begin{equation}
u+ i\phi_k=\lambda/2+ i\pi\veps^*/2 \qquad k=1,2,\dots,n(\alpha,\beta).
\end{equation}
Hence
\begin{equation}
\exp[-\sigma^{(\alpha,\beta)}]=
\left[\frac
{\thfr(0,|t|^\nu)}
{\thfr(\pi/2,|t|^\nu)}\right]^{n(\alpha,\beta)}
= k'(|t|^\nu)^{n(\alpha,\beta)/2}
\end{equation}
where $k'$ is the conjugate elliptic modulus
\begin{equation}
k'(p)=\prod_{n=1}^\infty\round{\frac{1-p^{2n-1}}{1+p^{2n-1}}}^4.
\end{equation}
It follows that the corresponding
critical exponent is given by
\begin{equation}
\sigma^{(\alpha,\beta)}\sim |t|^\mu \qquad
\mu=\nu=\begin{cases}
(L+1)/4 & \text{regime~III}\\
(L+1)/2 & \text{regime~IV.}
\end{cases}
\end{equation}

Notice that the phases are linearly ordered and that the only way to get from
one phase to another is to pass through any intermediate phases. So let us
define $\sigma^{(\alpha,\beta)}=0$ for the non-ground state phases. Similarly,
for $L$ odd in regime~IV, let us define $\sigma^{((L-1)/2^+,(L+1)/2^+)}=0$
etc.\ for the disordered phase.
Then the interfacial tensions are additive
and it follows, for example, that if $a<b$,
\begin{equation}
\sigma^{(a^+,b^+)}=\sum_{k=1}^{b-a}\sigma^{((a+k-1)^+,(a+k)^+)}.
\end{equation}

\section{Correlation lengths}

Let $\varphi(a)$ be a given function of the height $a$
and let $a_0$ and $a_\ell$ be two heights in the same column
separated by an even number $\ell$ of lattice spacings.
Then, at large distances, the (truncated) pair correlation function decays
exponentially with a correlation length $\xi$ given by
\begin{equation}
\langle\varphi(a_0)\varphi(a_\ell)\rangle-\langle\varphi(a_0)\rangle
\langle\varphi(a_\ell)\rangle\sim \exp(-\ell\xi^{-1}).
\end{equation}
By standard row transfer matrix arguments~\cite{BaxterBook} the correlation
length can be calculated from
\begin{equation}
-\xi^{-1}=\lim_{\ell\to\infty}\lim_{N\to\infty} \ell^{-1}
\ln \sum_{j\ne\smax} c_j \Lambda_j(u)^{\ell/2}
\label{corrlength}
\end{equation}
where the $\Lambda_j(u)$ are eigenvalues of
\begin{equation}
\Lambold(u)=\frac{\Dm^{(\alpha,\alpha)}(u)}{D^{(\alpha,\alpha)}_{\smax}(u)}
\end{equation}
and the $c_j$ denotes $u$-independent matrix elements.  The sum excludes the
maximal eigenvalue. Again the bulk and surface terms cancel out of the
ratio $\Lambold(u)$.
{}From the inversion relation and crossing, it follows that the eigenvalues
$\Lambda(u)$ again satisfy the simple functional equations
\begin{equation}
\Lambda(u)\Lambda(u+\lambda)=1 \qquad \Lambda(u)=\Lambda(\lambda-u)
\end{equation}
subject to the previous periodicity~\eqref{funcperiod}.

In the thermodynamic limit the sum in \eqref{corrlength} is dominated
by the first band of
$(2-\delta_{\alpha,1^+}-\delta_{\alpha,L^-}) \binom{N/2}{2}$ eigenvalues.
In this band the eigenvalues $\Lambda_j(u)$ are analytic
in an open strip
containing $0\le\Re(u)\le\lambda$, with two pairs of zeros at
\begin{equation}
u=\lambda/2\pm i\phi_1 \qquad u=\lambda/2\pm i\phi_2.
\end{equation}
The required solution of the inversion relation is therefore
\begin{equation}
\Lambda(u)=\prod_{k=1}^2\Phi(u+i\phi_k)\Phi(u-i\phi_k)
\end{equation}
where $\Phi(u)$ is given by \eqref{PhiIII}.
Passing to the thermodynamic limit for real $u$ in the interval
$0<u<\lambda$ we obtain
\begin{equation}
\lim_{N\to\infty}\sum_{j\ne \smax}c_j\Lambda_j(u)^{\ell/2}\sim
\int_0^{\frac{\pi\eps^*}{2}}\!\!\int_0^{\frac{\pi\veps^*}{2}}c(\phi_1,\phi_2)
\prod_{k=1}^2 |\Phi(u+i\phi_k)|^\ell\, d\phi_1d\phi_2.
\end{equation}
Carrying out the saddle point analysis as before gives
\begin{equation}
\exp(-\xi^{-1})=k'(|t|^\nu)
\end{equation}
where $k'$ is the conjugate elliptic modulus and hence
\begin{equation}
\xi\sim |t|^{-\nu} \qquad \nu=
\begin{cases}
(L+1)/4 &\text{regime~III}\\
(L+1)/2 &\text{regime~IV.}
\end{cases}
\end{equation}

It can be seen that, after allowing for differences in formulation,
all of the above results for the interfacial tensions
and correlation lengths agree with the known results in the case of the
Ising model~\cite{BaxterBook} ($L=3$) and the interacting hard square
model~\cite{BaxP} ($L=4$).
Notice also that we have the simple relation
\begin{equation}
\xi^{-1}=2\sigma^{(1^+,2^+)}.
\end{equation}
The factor of $2$ is easily understood because in the saddle point
integrals the contribution to
the inverse correlation length involves two domain walls, whereas the
contribution to this interfacial tension derives from just one such domain
wall.

Finally, we note that
in regime~III the critical exponents $\alpha$, $\als$, $\mu$ and $\nu$ satisfy
the scaling relations~\cite{BaxterBook,Binder,Diehl}
\begin{equation}
2-\alpha=\mu+\nu=2\nu \qquad \als=\alpha+\nu.
\end{equation}
However, these scaling relations break down in regime~IV, since the regime~IV
line of exact solution approaches the tricritical point $t=0$ at a tangent to
the critical line~\cite{Huse83}.

\section{Discussion}

In this paper we have calculated the surface free energies, interfacial
tensions and correlation lengths of the ABF models in regimes~III and IV,
all by solving a relatively simple inversion relation. The methods are
general and can be applied to other solvable lattice models such as the
CSOS models and dilute lattice models.
The methods employed here could be applied to regimes~I and II with fixed
boundary conditions where the heights alternate along the boundary. However,
in regimes~I and II, the coexisting phases are selected out by a saw-tooth
variation in the ground state configuration~\cite{ABF84,Huse84} and
unfortunately these saw-tooth variations on the boundary cannot be handled
within our formalism at present. So, strictly speaking, the current methods
are best suited to systems that exhibit at most two independent
sublattices in the structure of their ground states.

Of course, the correlation length along the strip is a bulk property and
not a surface property and so its calculation should not necessitate the
introduction of a boundary. Accordingly, the generalized inversion relation
method used here to obtain the correlation length can be applied in the case
of periodic boundary conditions simply by working with a double-row transfer
matrix
\begin{equation}
\Dm(u)=\Tm(u)\Tm(\lambda-u).
\end{equation}
Since the correlation length is
independent of the spectral parameter $u$, and since $\Dm(u)$ and $\Tm(u)$
yield the same correlation length at the isotropic point $u=\lam/2$, the two
transfer matrices must lead to the same correlation length for all values of
$u$.

\section*{Acknowledgements}

This work was done while PAP was visiting Bonn and Amsterdam Universities.
It is a pleasure to thank Vladimir Rittenberg and Bernard Nienhuis for their
kind hospitality. We also thank Roger Behrend for his collaboration on previous
related work, and Ole Warnaar for his useful suggestions.
This research is supported by the Australian Research Council.


\begin{thebibliography}{00}

\bibitem{BaxterBook} R. J. Baxter, ``Exactly Solved Models in Statistical
Mechanics''.  Academic Press, London, 1982.

\bibitem{Chered} I. V. Cherednik, Teor. Mat. Fiz. {\bf 61} (1984) 35.

\bibitem{Sk88} E. K. Sklyanin, J. Phys.\ A \textbf{21} (1988) 2375.

\bibitem{BPO95} R. E. Behrend, P. A. Pearce and D. L. O'Brien, J. Stat.\ Phys.
\textbf{84} (1996) 1.

\bibitem{Kul95} P. P. Kulish, \textit{Yang-Baxter equation and reflection
equations in integrable models}, hep-th/9507070.

\bibitem{AK95} C. Ahn and W. M. Koo, \textit{Boundary Yang-Baxter equation in
the RSOS representation}, Preprint EWHA-TH, SNUTP-95-080, hep-th/9508080.

\bibitem{OPB95} D. L. O'Brien, P. A. Pearce and R. E. Behrend, \textit{Surface
Free Energies and Surface Critical Behaviour of the ABF Models with Fixed
Boundaries}, cond-mat/9511081, to be published in the
proceedings of \emph{Statistical Models, Yang-Baxter Equation and Related
Topics}, Tianjin, China, 1995.

\bibitem{ZB95} Y. K. Zhou and M. T. Batchelor, \textit{Surface Critical
Phenomena
in Inter\-action-Round-a-Face Models},  ANU preprint MRR-070-95,
cond-mat/9511008.

\bibitem{BZ95} M. T. Batchelor and Y. K. Zhou, Phys.\ Rev.\ Lett.\ \textbf{76}
(1996) 14; 2826.

\bibitem{BFZ96} M. T. Batchelor, V. Fridkin and Y. K. Zhou, J. Phys.\ A
\textbf{29} (1996) L61.

\bibitem{ZB96} Y. K. Zhou and M. T. Batchelor, J. Phys.\ A \textbf{29}
(1996) 1987.

\bibitem{Stroganov} Yu. G. Stroganov, Phys.\ Lett.\ \textbf{74A} (1979) 116.

\bibitem{Bax82} R. J. Baxter, J. Stat.\ Phys.\ \textbf{28} (1982) 1.

\bibitem{ABF84} G. E. Andrews, R. J. Baxter and P. J. Forrester, J.
Stat.\ Phys.\ \textbf{35} (1984) 193.

\bibitem{Huse84} D. A. Huse, Phys.\ Rev.\ B. \textbf{30} (1984) 3908.

\bibitem{BaxP} R. J. Baxter and P. A. Pearce, J. Phys.\ A \textbf{15} (1982)
897; J. Phys.\ A \textbf{16} (1983) 2239.

\bibitem{Binder} K. Binder, \textit{in} ``Phase Transitions and Critical
Phenomena'', Volume 8, (C.~Domb and J. L. Lebowitz, eds).  Academic
Press, London, 1983.

\bibitem{Diehl} H. W. Diehl, \textit{in} ``Phase Transitions and Critical
Phenomena'', Volume 10, (C.~Domb and J. L. Lebowitz, eds).  Academic
Press, London, 1986.

\bibitem{Huse83} D. A. Huse, J. Phys.\ A \textbf{16} (1983) 4357.

\end{thebibliography}
\end{document}